\documentclass[aps,prl,twocolumn]{revtex4-1}

\usepackage{url}
\usepackage{graphicx}
\usepackage{amsmath}

\begin{document}
%\markboth{Ho et al.}{Modeling cell size regulation}
\title[Modeling cell size regulation]{Modeling cell size regulation: From single-cell level statistics
to molecular mechanisms and population level effects}
\author{Po-Yi Ho}
\author{Jie Lin}
\author{Ariel Amir*}
\affiliation{John A. Paulson School of Engineering and Applied Sciences, Harvard University, Cambridge, MA, USA, 02138}
\email{arielamir@seas.harvard.edu}
%\author{Po-Yi Ho, Jie Lin, and Ariel Amir*
%\affil{John A. Paulson School of Engineering and Applied Sciences, Harvard University, Cambridge, MA, USA, 02138}
%\affil{arielamir@seas.harvard.edu}}

\begin{abstract}
Most microorganisms regulate their cell size. We review here some
of the mathematical formulations of the problem of cell size regulation.
We focus on coarse-grained stochastic models and the statistics they
generate. We review the biologically relevant insights obtained from
these models. We then describe cell cycle regulation and their molecular
implementations, protein number regulation, and population growth,
all in relation to size regulation. Finally, we discuss several future
directions for developing understanding beyond phenomenological models
of cell size regulation.
\end{abstract}

%\begin{keywords}
%cell size regulation, coarse-grained stochastic
%models, microbial cell cycle, molecular mechanisms, single-cell variability,
%population growth
%\end{keywords}

\maketitle

%\tableofcontents

\section{Introduction}

Most microorganisms regulate their cell size, as evidenced by their
narrow cell size distributions. In particular, all known species of
bacteria have cell size distributions with small coefficient of variations
(CV, standard deviation divided by the mean), which can be as low
as 0.1 \cite{koch}. For cells that grow exponentially, a small CV
for size implies a small CV for interdivision times. However, a small
CV for interdivision times is not sufficient to regulate cell size,
as we will show that a simple ``timer'' strategy cannot regulate
cell size in face of fluctuations. Cells must therefore have a way
to effectively measure size.

The physiological implications of cell size remain under debate. In
the context of bacteria, this is discussed in detail in a recent,
excellent review \cite{huang}, which also stresses the intimate connection
between the problem of cell size regulation and that of cell cycle
regulation. For instance, cell division, which mechanistically determines
cell size, is coupled to DNA replication. In this review, we will
not focus on the rich biology behind this problem, but instead will
elaborate on the various phenomenological models developed to study
this problem over the last several decades. These are typically coarse-grained
models, which consider the cell as a whole and describe cell volume
at various stages of the cell cycle. They often seek to capture the
statistics of the random process underlying cell size regulation.
For example, what is the relation between cell size and interdivision
time, and what distributions characterize the fluctuations in these
variables? 

A devil's advocate or a biologist may ask why one would care about
these questions. Quantitatively describing the distributions and finding
scaling relations between variables are worthy goals from a physicist's
statistical mechanical point of view, but can such phenomenological
modeling shed light on the biology? Three distinct examples support
that the answer to this question is affirmative.

First, for several bacterial model species, including \emph{E. coli
}and \emph{B. subtilis}, cell volume scales exponentially with growth
rate, and proportionally with, loosely speaking, chromosome copy number
\cite{smk,levin,jun}. The scaling constant is in fact equal to the
time from the initiation of DNA replication to cell division. It was
shown fifty years ago that this observation can be rationalized within
a model in which the regulation of cell size does not occur via controlling
the timing of cell divisions, but rather via controlling the timing
of the initiation of DNA replication \cite{donachie}. In this way,
a quantitative pattern on the phenomenological level, with the aid
of mathematical modeling, led to an important insight regarding bacterial
physiology. The same empirical observation helped to address whether
cell size regulation occurs over cell volume, surface area, or other
dimensions. Experiments in rod-shaped bacteria often measure cell
length, which cannot distinguish between these possibilities since
cell width in these bacteria is very narrowly distributed (CV $<0.05$)
\cite{jun}. As a result, in addition to cell volume, both cell surface
area and length have been proposed to set cell size \cite{theriot,jw}.
However, recent experiments in \emph{E. coli} showed that the same
scaling relation holds, but only for cell volume and not surface area
or width, under genetic perturbations to cell dimensions \cite{liu}.
This result supports that volume is the key phenomenological variable
controlling cell size. Below, we use the term cell size for generality
while keeping the above discussion in mind.

Second, a naive proposal for cell size regulation is a timer strategy,
in which cells control the timing of their cell cycles so that, on
average, cell size doubles from birth to division. However, it can
be shown by theoretical arguments alone that this mode of regulation
is incompatible with the small CVs of cell size distributions if cell
volume grows exponentially in time at the single-cell level, as seen
in experiments \cite{godin}. This is because the cumulative effect
of noise will cause the variance in cell size to diverge. Explicitly,
consider exponentially growing cells with a constant growth rate $\lambda$
and stochastic interdivision time $t_{d}$. A cell born at size $v_{b}$
will generate a progeny of size $v_{b}'=v_{b}e^{\lambda t_{d}}/2$,
assuming perfect symmetric division. Let $x=\ln\left(v_{b}/v_{0}\right)$
be the log-size, where $v_{0}$ is a constant that sets the mean cell
size, and $x'$ the log-size at the next generation, then 
\begin{equation}
x'=x+\lambda t_{d}-\ln2.\label{eq:rw}
\end{equation}
Uncorrelated fluctuations in $t_{d}$ will then lead to a random walk
in log-sizes with fluctuations accumulating as the square root of
the number of divisions. Thus, the cell size distribution in a growing
population will not reach stationarity via a timer strategy (Fig.
\ref{fig:data}a), and a different strategy is needed to achieve narrow
distributions (Fig. \ref{fig:data}b). Note that without fluctuations,
Eq. \ref{eq:rw} becomes $x'=x$ if $t_{d}=\ln2/\lambda$, so that
cell size is maintained. This example therefore shows that it is necessary
to introduce stochasticity to models of cell size regulation as the
failure of a timer strategy cannot be revealed otherwise.

As a third example, some of us recently investigated the properties
of the resulting size regulation strategy from models of molecular
mechanisms that do not specify the identity of the molecular players,
but nonetheless propose concrete molecular network architectures.
Two models were considered, one proposing that cell division is triggered
by the accumulation to a threshold number of an initiator protein
\cite{sompayrac}, and another that the dilution of an inhibitor triggers
an event in cell cycle progression \cite{fantes}. It was shown theoretically
that in the context of budding yeast \emph{S. cerevisiae}, both of
these seemingly reasonable size regulation strategies fail to regulate
cell size in the case of symmetric division \cite{barber}. While
there could be other explanations, this appears to be a strong constraint
that may have contributed to the evolution of asymmetric division
in budding yeast.

These very different examples show how phenomenological modeling,
in combination with single-cell or bulk-level level experiments quantifying
cell growth, can lead to biologically relevant conclusions and constrain
biological mechanisms. Furthermore, this approach allows to construct
a theoretical \textquotedblleft phase diagram\textquotedblright{}
(e.g. \cite{marantan}), showing not where biology lies, but where
biology may exist. In this vein, we proceed with the following aphorism
in mind, ``All models are wrong; some models are useful.'' \cite{boxQuote}

In this review, we describe various existing phenomenological models
for cell size regulation, some dating decades back and many very recent,
and also present several novel results. First, we introduce discrete
stochastic maps (DSM) to model cell size regulation and the various,
approximate methods of solving for the distributions and correlations
they generate. We discuss the connection between the problem of cell
size regulation and that of diffusion in a confining potential and
autoregressive modeling in time-series analysis. We systematically
show, for the first time to our knowledge, that cell size regulation
in \emph{E. coli} can be approximated well by a stochastic model where
the cell size at the next generation depends only on the cell size
at the present generation. Next, we review continuous rate models
and their mapping to DSMs. We then review recent works that analyzed
DSMs at higher precision. They revealed that while the qualitative
stability regions can be found via approximate methods, detailed statistical
results are more nuanced, and specifically, power-law tails may often
be generated by DSMs. Finally, we review recent results beyond the
phenomenological level, including molecular implementations of different
strategies for cell size regulation, the problem of protein number
regulation, and the effects of cell size regulation on population
growth.

\section{Models for cell size regulation}

To resolve the problem of an unconfined random walk in log-sizes in
Eq. \ref{eq:rw}, feedback must be introduced so that larger cells
divide sooner than average. Some intuition for the problem of cell
size regulation can be gained by considering the familiar scenario
of overdamped Brownian motion in a confining potential. This scenario
can be described by the Langevin equation describing the dynamics
of position $x$ \cite{gardiner},
\begin{equation}
\frac{dx}{dt}=-\frac{1}{\gamma}V'\left(x\right)+\sigma\xi.\label{eq:langevin}
\end{equation}
Here, $\gamma$ is a drag coefficient that relates the force to the
velocity in the overdamped limit, $V\left(x\right)$ is the confining
potential, and $\sigma$ is the magnitude of the fluctuations described
by the stochastic variable $\xi$, which has correlations $\left\langle \xi\left(t'\right)\xi\left(t'+t\right)\right\rangle =\delta\left(t\right)$.
In this review, $\left\langle \cdot\right\rangle $ denotes the ensemble
average. In the absence of a potential $V\left(x\right)=0$, Eq. \ref{eq:langevin}
reduces to unconfined diffusion, whose hallmark is the linear dependence
of the mean-squared-displacement $\left\langle x^{2}\right\rangle $
on time. In a quadratic potential $V\left(x\right)=kx^{2}/2$, Eq.
\ref{eq:langevin} corresponds to diffusion confined by a linear restoring
force. In this case, Eq. \ref{eq:langevin} is known as an Ornstein-Uhlenbeck
(OU) process and is useful in describing a plethora of physical phenomena.
It can be written as
\begin{equation}
\frac{dx}{dt}=-\frac{k}{\gamma}x+\sigma\xi,\label{eq:ou}
\end{equation}
where $k$ is the strength of the restoring force.

The probability density $p\left(x,t\right)$ corresponding to Eq.
\ref{eq:ou} satisfies the Fokker-Planck equation that describes its
temporal dynamics \cite{gardiner},
\begin{equation}
\frac{\partial p}{\partial t}=\frac{k}{\gamma}\frac{\partial}{\partial x}\left(xp\right)+\frac{\sigma^{2}}{2}\frac{\partial^{2}p}{\partial x^{2}}.
\end{equation}
The stationary $\partial p/\partial t=0$ solution is a Gaussian distribution
\begin{equation}
p\left(x\right)=\sqrt{\frac{k}{\pi\gamma\sigma^{2}}}\exp\left(-\frac{kx^{2}}{\gamma\sigma^{2}}\right).\label{eq:gaussian}
\end{equation}
Indeed, Eq. \ref{eq:gaussian} is equal to the Boltzmann distribution
$p\left(x\right)\propto\exp\left(-V\left(x\right)/k_{B}T\right)$,
where $k_{B}$ is the Boltzmann constant and $T$ is the temperature,
since $\sigma^{2}=2D=2k_{B}T/\gamma$ by the Einstein relation for
the diffusion coefficient $D$. As $k\to0$, the strength of the confining
potential weakens. At $k=0$, the variance of $x$ diverges. However,
for any $k>0$, the variance of $x$ will be finite. The autocovariance
$\left\langle x\left(t'\right)x\left(t'+t\right)\right\rangle $ can
be obtained via integration of Eq. \ref{eq:ou}. At stationarity,
the autocovariance is exponentially decaying \cite{gardiner},
\begin{equation}
\left\langle x\left(t'\right)x\left(t'+t\right)\right\rangle =\frac{\gamma\sigma^{2}}{2k}\exp\left(-\frac{k}{\gamma}\left|t\right|\right).\label{eq:expdecay}
\end{equation}

The familiar example of an OU process turns out to be similar to the
problem of cell size regulation, but with the variable $x$ now representing
cell size. We now review the formulation of the problem of cell size
regulation as a discrete analogue of an OU process.

\subsection{Discrete stochastic maps\label{subsec:dsm}}

\begin{figure*}
\noindent \begin{centering}
\includegraphics[width=6in]{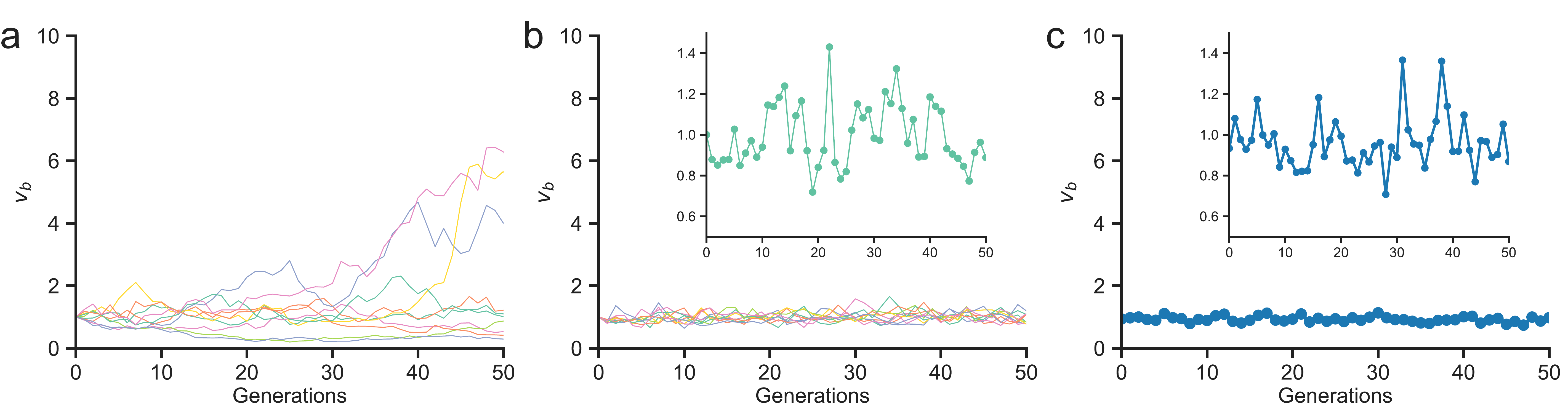}
\par\end{centering}
\caption{\label{fig:data} Cell size at birth in simulations (a,b) and in experiments
(c) as a discrete time-series, with insets showing a zoomed in view
of one particular trial of simulation (b) and the data (c). (a,b)
Multiple trials (different colors) of numerical simulations of the
DSM in Eq. \ref{eq:DSM} with the simple, one-line pseudo-code: \emph{$v_{i+1}=\left(2\left(1-\alpha\right)v_{i}+v_{0}\right)2^{\xi_{t}}/2$},
where $v_{i}$ denote cell size at birth in the $i$-th generation,
$\xi_{t}$ is a normally distributed random variable with zero mean
and variance $\sigma_{t}^{2}$, and $v_{0}$ is a constant that sets
the mean cell size. Here, $\sigma_{t}=0.22$ and $\left\langle v_{b}\right\rangle =1$.
(a) $\alpha=0$ leads to unconfined diffusion and a divergent distribution.
(b) Any $0<\alpha<2$, here $\alpha=0.5$, has the necessary feedback
to achieve a stationary distribution. (c) Data from Ref. \cite{youData}.
$v_{b}$ in this case represents cell length at birth and is normalized
so that $\left\langle v_{b}\right\rangle =1$.}
\end{figure*}

Fig. \ref{fig:data}c shows single-cell data obtained via microfluidic
devices that trap single cells in micro-channels to allow measurements
of physiological properties such as cell size for many generations
\cite{youData,junWang,taheriaraghiRev}. The problem of cell size
regulation may be investigated initially by considering only division
events. The data in this case consist of cell size at birth, division,
and interdivision time over many generations. What are the distribution
and correlations of cell size at birth and division, and what size
regulation strategies lead to such statistics?

At a phenomenological, coarse-grained level, a size regulation strategy
can be specified as a map that takes cell size at birth $v_{b}$ to
a targeted cell size at division $v_{a}$ with a deterministic strategy
$f\left(v_{b}\right)$ \cite{amir},
\begin{equation}
v_{a}=f\left(v_{b}\right).\label{eq:fvb}
\end{equation}
In face of biological stochasticity, the actual cell size at division
$v_{d}$ is $v_{a}$ subject to some coarse-grained noise term. For
example, the noise term can be size-additive, so that $v_{d}=v_{a}+\xi_{v}$,
where $\xi_{v}$ is uncorrelated between generations. The noise term
can also be time-additive. In this case, the stochastic interdivision
time $t_{d}$ can be written as $t_{d}=t_{a}+\xi_{t}$, where $\xi_{t}$
is the noise term. The deterministic component $t_{a}$ can be determined
by assuming a constant exponential growth rate $\lambda$. The deterministic
size regulation strategy in Eq. \ref{eq:fvb} then leads to 
\begin{equation}
t_{a}=\ln\left(f\left(v_{b}\right)/v_{b}\right)/\lambda.\label{eq:ta}
\end{equation}
In the case of time-additive noise, if division is perfectly symmetric
so that $v_{b}'=v_{d}/2$, then the cell size at birth at the next
generation is
\begin{equation}
v_{b}'=f\left(v_{b}\right)e^{\lambda\xi_{t}}/2.\label{eq:DSM}
\end{equation}
The two forms of noise lead to distributions of different shapes.
Experiments have shown that distributions of cell sizes at birth are
skewed and can be approximated as a log-normal but that interdivision
time distributions can be approximated as normal \cite{jw,elf}. These
are consistent with a normally distributed time-additive noise, which
we use below.

\subsection{Approximate solution via first order expansion\label{subsec:firstorder}}

The DSM described in Section \ref{subsec:dsm} is in general difficult
to solve for an arbitrary size regulation strategy $f\left(v_{b}\right)$.
One method makes the approximation to focus on the behavior of $f$
near the mean size $\left\langle v_{b}\right\rangle $, since the
size distribution has a small CV. A size regulation strategy can be
linearized by expanding about $\left\langle v_{b}\right\rangle $,
$f\left(v_{b}\right)\approx f\left(\left\langle v_{b}\right\rangle \right)+f'\left(\left\langle v_{b}\right\rangle \right)\left(v_{b}-\left\langle v_{b}\right\rangle \right).$
In this approximation, all regulation strategies that agree to first
order will lead to similar distributions near $\left\langle v_{b}\right\rangle $.
The following is a convenient choice \cite{amir},
\begin{equation}
f\left(v_{b}\right)=2v_{b}^{1-\alpha}v_{0}^{\alpha},\label{eq:fchoice}
\end{equation}
where $v_{0}$ is an arbitrary constant. As shown below, $\left\langle v_{b}\right\rangle \approx v_{0}$,
and hence, the slope has value $f'\left(\left\langle v_{b}\right\rangle \right)=2\left(1-\alpha\right)$.
The value of $\alpha$ therefore determines the strength of regulation.
$\alpha=1$ corresponds to the strongest regulation, a ``sizer''
strategy where cells attempt to divide upon reaching $f\left(v_{b}\right)=2v_{0}$.
$\alpha=0$ represents no regulation and corresponds to the timer
strategy where cells attempt to divide upon reaching $f\left(v_{b}\right)=2v_{b}$.
Recent works have shown that the statistics of cell size can be generated
by a regulation strength $\alpha=1/2$ that is between the two extremes
\cite{jun,jw,elf,junReview}. In this case, the slope has value $f'\left(\left\langle v_{b}\right\rangle \right)=1$
and so is an approximation to the ``adder'' strategy (also known
as the incremental model \cite{voorn,koppes}) where cells attempt
to divide upon reaching $f\left(v_{b}\right)=v_{b}+v_{0}$. Several
microorganisms in all three domains of life have been shown to approximately
follow an adder strategy, or less prescriptively, to exhibit adder
correlations. We discuss the prevalence of adder correlations later.

Let $x=\ln\left(v_{b}/v_{0}\right)$ be the log-size and $x'$ denote
$x$ at the next generation. Eqs. \ref{eq:DSM}-\ref{eq:fchoice}
then lead to the simple stochastic equation

\begin{equation}
x'=\left(1-\alpha\right)x+\lambda\xi_{t}.\label{eq:xprime}
\end{equation}
At the $n$-th generation,
\begin{equation}
x_{n}=\left(1-\alpha\right)^{n}x_{0}+\sum_{j=0}^{n-1}\left(1-\alpha\right)^{n-1-j}\lambda\xi_{t}^{\left(j\right)},\label{eq:series}
\end{equation}
where $x_{i}$ and $\xi_{t}^{\left(i\right)}$ respectively denote
the value of $x$ and $\xi_{t}$ at the $i$-th generation. The first
term approaches zero as $n\to\infty$ if $0<\alpha<2$. If $\xi_{t}$
is normally distributed with variance $\sigma_{t}^{2}$, then the
variance $\sigma_{x}^{2}$ of $x$ will be the sum of the variances
in the series in the second term. The geometric series converges for
$0<\alpha<2$, and can readily be evaluated to give the variance $\sigma_{x}^{2}$
as
\begin{equation}
\sigma_{x}^{2}=\frac{\lambda^{2}\sigma_{t}^{2}}{\alpha\left(2-\alpha\right)}.\label{eq:sigx2}
\end{equation}
Furthermore, since $x_{n}$ is a sum of normal variables, it will
also be normally distributed. If $\alpha\leq0$ or $\alpha\geq2$,
the sum of the series diverges, and hence there is no stationary distribution.
The case $\alpha=0$ produces unconfined diffusion and is analogous
to the case where the strength of the restoring force is zero ($k=0$)
in an OU process, as seen in Eq. \ref{eq:gaussian}. Fig. \ref{fig:data}ab
demonstrates the difference between time-series generated by $\alpha=0$
and by $0<\alpha<2$. The variance of $t_{d}$ can be obtained similarly.

It is not obvious a priori\emph{ }whether the widths of the distributions
of interdivision time and cell size are related. It turns out that
the two CVs (denoted by $CV\left(\cdot\right)$) are related by a
dimensionless quantity \cite{amir}. The log-size is related to the
actual size by $x=\ln\left(v_{b}/v_{0}\right)\equiv\ln\left(1+\delta v_{b}\right)$.
Since $\delta v_{b}=v_{b}/v_{0}-1$ is small, $x\approx\delta v_{b}=v_{b}/v_{0}-1$.
Therefore, $CV\left(v_{b}\right)\approx\sigma_{x}$. Calculating $CV\left(t_{d}\right)$
in a similar manner leads to
\begin{equation}
\frac{CV\left(v_{b}\right)}{CV\left(t_{d}\right)}\approx\frac{\ln2}{\sqrt{2\alpha}}.\label{eq:cvratio}
\end{equation}
Eq. \ref{eq:cvratio} allows to extract the parameter $\alpha$ from
CVs that can be accurately measured. Since $x$ and $t_{d}$ are both
distributed normally, the model predicts that these distributions
can be collapsed after normalizing by the mean and scaling according
to Eq. \ref{eq:cvratio}, as seen in experiments \cite{jun,soifer}.

The Pearson correlation coefficients (CC) between two variables (denoted
by $C\left(\cdot,\cdot\right)$) can also be obtained. Since CCs are
not affected by addition or multiplication by a constant, $v_{b}$
can be replaced by $x$ in the following calculations. The CC between
cell size at birth of a mother cell and that of the daughter cell
is therefore \cite{amir}
\begin{equation}
C\left(v_{b},v_{b}'\right)=C\left(x,x'\right)=\frac{\left\langle xx'\right\rangle -\left\langle x\right\rangle ^{2}}{\sigma_{x}^{2}}.
\end{equation}
Substituting in Eq. \ref{eq:xprime},
\begin{equation}
C\left(v_{b},v_{b}'\right)=1-\alpha.\label{eq:cvbvd}
\end{equation}
Importantly, the value of $\alpha\approx1/2$ extracted via the ratio
of CVs in Eq. \ref{eq:cvratio} also predicts the CC between size
at birth and at division, in agreement with experiments \cite{jun,jw,koppes80}.

Similarly, using Eq. \ref{eq:ta} and \ref{eq:fchoice}, the interdivision
time can be written as
\begin{equation}
t_{d}=\frac{\ln2-\alpha x}{\lambda}+\xi_{t}.
\end{equation}
The CC between the interdivision times of a mother-daughter pair can
then be shown to be \cite{jun,lin}
\begin{equation}
C\left(t_{d},t_{d}'\right)=-\alpha/2.\label{eq:ctdtd}
\end{equation}
That this CC is non-zero has implications for the population growth
rate, which we review later.

\begin{figure*}
\noindent \begin{centering}
\includegraphics[width=6in]{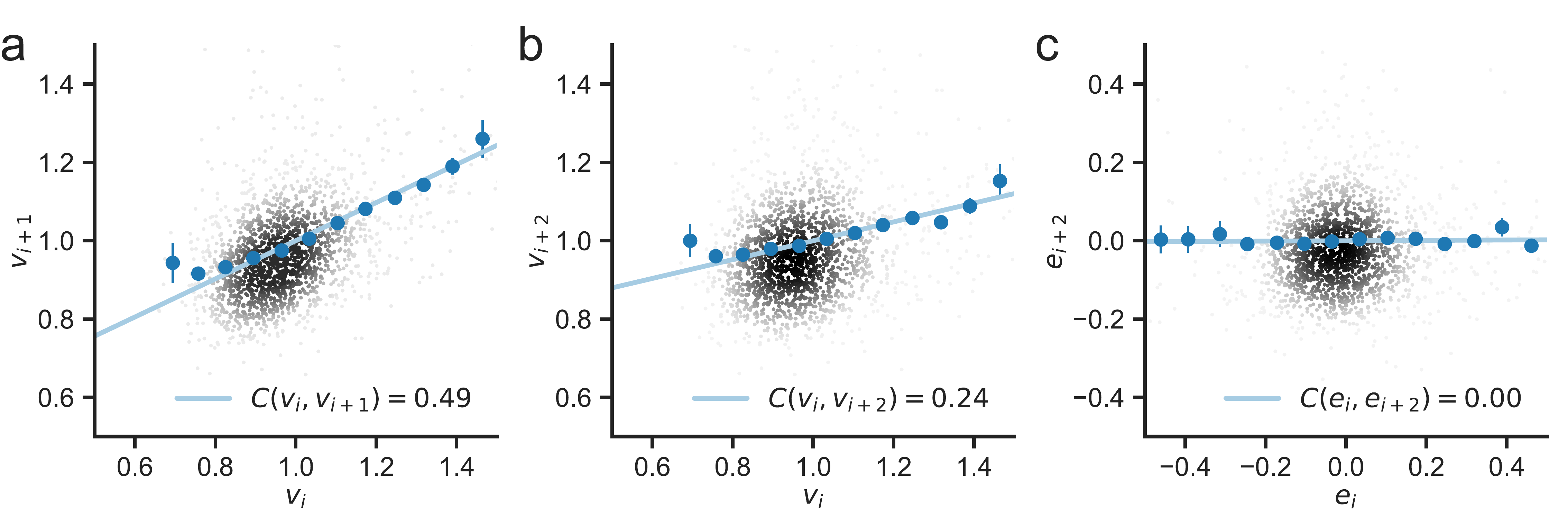}
\par\end{centering}
\caption{\label{fig:analysis} Cell size at birth in \emph{E. coli }can be
described by an AR(1) model. Data is the same as that in Fig. \ref{fig:data}
\cite{youData}. (a-b) Single-cell data of cell size at birth $v_{i}$
at the $i$-th generation and the resulting CCs between parent and
children (a) and parent and grand-children (b). (c) The residuals
$e_{i}$ after linear regression with $v_{i+1}$ and the resulting
CC, $C\left(e_{i},e_{i+2}\right)$. See text for details. Gray dots
show data, with color representing the density of points. Blue circles
show average values binned according to values on the $x$-axis. Error
bars show one SEM. Blue lines show the best linear regression of raw
data.}
\end{figure*}

\subsection{Autoregressive models and extensions to incorporate biological details}

Eq. \ref{eq:xprime}, obtained after linearization of the generically
nonlinear DSM Eq. \ref{eq:DSM}, is mathematically known as an autoregressive
(AR) model, often used in time-series analysis and economics forecasting
\cite{priestley}. An AR model of order $m$ (denoted by AR($m$))
takes the form $x_{i}=b+\sum_{j=1}^{m}c_{j}x_{i-j}+\xi_{i}$, where
$b$ and $c_{j}$ are constants and $\xi_{i}$ is a noise term uncorrelated
for different $i$. The model describes how previous values of the
stochastic variable $x$ influence linearly the next value. Eq. \ref{eq:xprime}
is an AR(1) model, which is also a discrete analogue of an OU process.

In the problem of cell size regulation, it is not obvious a priori\emph{
}if an AR(1) model is sufficient to describe data. One method to determine
the appropriate order of an AR model is to investigate the partial
correlation coefficients. The partial CC between $x_{i}$ and $x_{i+2}$
given an intermediate variable $x_{i+1}$ is defined as the CC between
the residuals $e_{i}=x_{i}-\hat{x}_{i}$ and $e_{i+2}=x_{i+2}-\hat{x}_{i+2}$.
$\hat{x}_{i}=p_{1}x_{i+1}+p_{0}$ and $\hat{x}_{i+2}=q_{1}x_{i+1}+q_{0}$
denote the predicted value after linear regression with $x_{i+1}$
to determine the coefficients. The resulting partial CC between $x_{i}$
and $x_{i+2}$, with the intermediate variable $x_{i+1}$, is \cite{priestley}
\begin{equation}
\begin{split}
C_{x_{i+1}}\left(x_{i},x_{i+2}\right)=C\left(e_{i},e_{i+2}\right)\\
=\frac{C\left(x_{i},x_{i+2}\right)-C\left(x_{i},x_{i+1}\right)C\left(x_{i+1},x_{i+2}\right)}{\sqrt{\left(1-C^{2}\left(x_{i},x_{i+1}\right)\right)\left(1-C^{2}\left(x_{i+1},x_{i+2}\right)\right)}}.
\end{split}
\end{equation}
In an AR(1) model, the CC $C\left(x_{i},x_{i+2}\right)$ is non-zero
because they are related via the intermediate variable $x_{i+1}$.
However, the partial CC $C_{x_{i+1}}\left(x_{i},x_{i+2}\right)$ removes
the effects of the intermediate variable and is zero. Experimentally
determined values of $C\left(v_{i},v_{i+1}\right)$ is as predicted
by Eq. \ref{eq:cvbvd} (Fig. \ref{fig:analysis}a) \cite{youData}.
In the same data set, $C\left(v_{i},v_{i+2}\right)$ is non-zero but
$C_{v_{i+1}}\left(v_{i},v_{i+2}\right)$ is zero (Fig. \ref{fig:analysis}bc).
Indeed, a vanishing $C_{v_{i+1}}\left(v_{i},v_{i+2}\right)$ implies
that $C\left(v_{i},v_{i+2}\right)=C^{2}\left(v_{i},v_{i+1}\right)$,
which is the case here. This novel check systematically shows that
cell size at birth in \emph{E. coli} can be described by an AR(1)
model. This result is a fortunate simplification, since for example,
in certain mammalian cells, the CCs in the interdivision times between
cousin cells ($C_{\textnormal{cc}}$) cannot be determined from those
between sister cells ($C_{\textnormal{ss}}$) and between mother-daughter
pairs ($C_{\textnormal{md}}$). Instead, experiments observe that
$C_{\textnormal{cc}}>C_{\textnormal{md }}$, contrary to the expected
relation $C_{\textnormal{cc}}=C_{\textnormal{md}}^{2}C_{\textnormal{ss}}$
in an AR(1) model \cite{balaban,balaban2}.

\begin{figure*}
\noindent \begin{centering}
\includegraphics[width=6in]{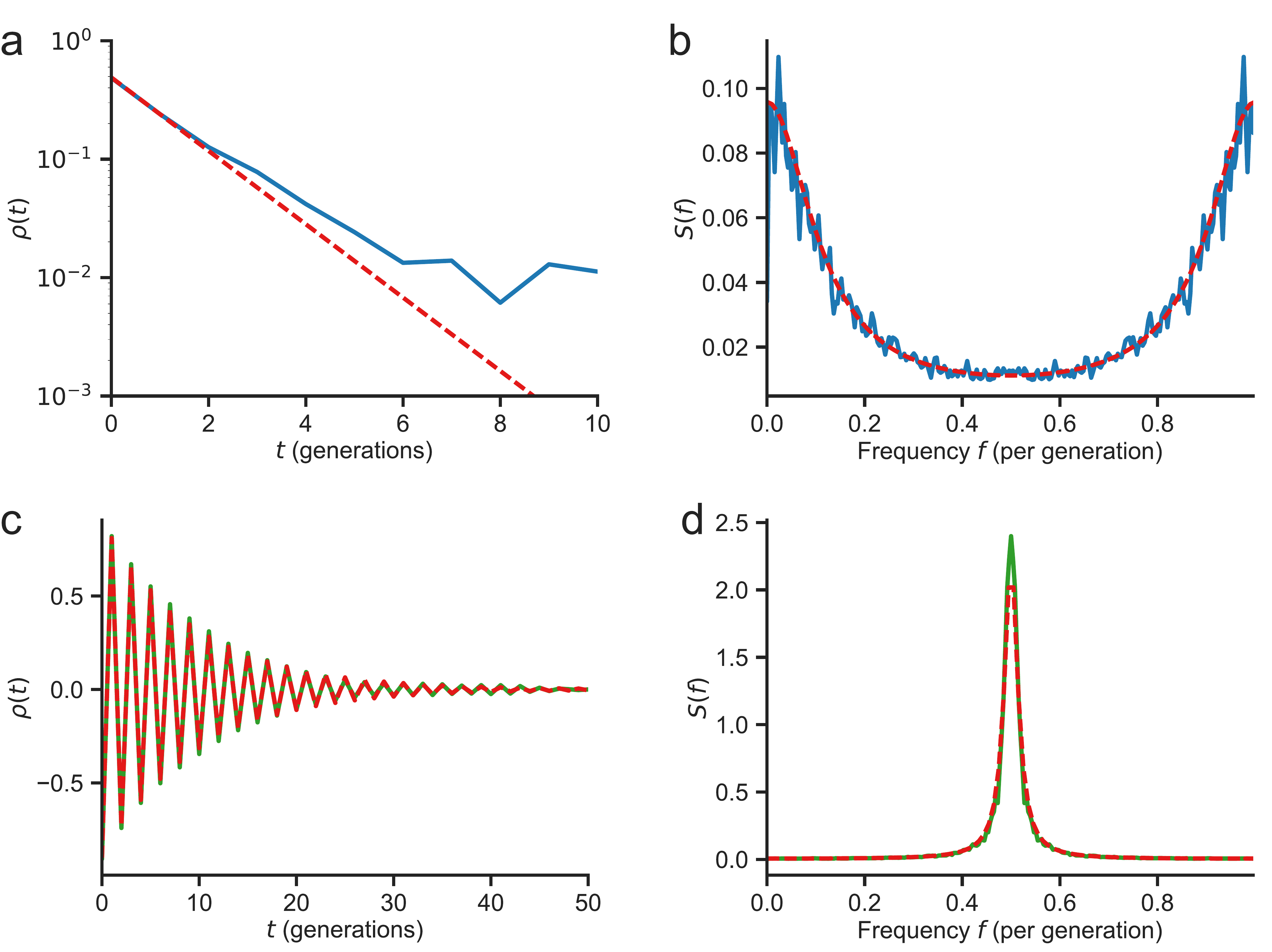}
\par\end{centering}
\caption{\label{fig:oscillations} Fluctuations versus oscillations. The ACF
$\rho\left(t\right)$ (a,c) and the PSD $S\left(f\right)$ (estimated
via the Welch method) (b,d) of cell size at birth in \emph{E. coli
}(a,b) and a simulated AR(1) model described by Eq. \ref{eq:xprime}
with $\alpha=1.9$ (c,d). Blue lines show experimentally determined
ACF and PSD from the same data set as that in Fig. \ref{fig:data}
\cite{youData}. Green lines show simulation results. Dashed red lines
show Eq. \ref{eq:acfMod} (a,c) and Eq. \ref{eq:sfMod} (b,d), for
$\alpha=0.49$ (a,b) and $\alpha=1.9$ (c,d).}
\end{figure*}

Extracting the regulation strength $\alpha$ via Eq. \ref{eq:cvbvd}
is analogous to estimating the parameters in AR models via the Yule-Walker
equations that relate theoretical values of the parameters to theoretical
values of the autocorrelation function (ACF) \cite{priestley}. The
ACF $\rho\left(t\right)$ is the CC between variables separated by
$t$ time points, 
\begin{equation}
\rho\left(t\right)=C\left(x_{i},x_{i+t}\right).
\end{equation}
As can be seen by Eq. \ref{eq:series}, the ACF for the AR(1) model
of Eq. \ref{eq:xprime} is simply
\begin{equation}
\rho\left(t\right)=\left(1-\alpha\right)^{\left|t\right|}.\label{eq:acfMod}
\end{equation}
In this case, the ACF decays exponentially as in an OU process as
seen in Eq. \ref{eq:expdecay}. The ACF of cell size at birth indeed
decays exponentially (Fig. \ref{fig:oscillations}a) \cite{you}.
Importantly, the estimated ACF is only meaningful after sufficient
averaging to eliminate spurious fluctuations. This can be done most
clearly by computing the power spectral density (PSD)
\begin{equation}
S\left(f\right)=\lim_{T\to\infty}\frac{1}{T}\left|\sum_{j=1}^{T}x_{j}e^{-i2\pi fj}\right|^{2},
\end{equation}
where $T$ is the total number of observations in the time-series
with data points $x_{j}$. The PSD can also be calculated as the Fourier
transform of the ACF according to the Wiener-Khinchin theorem. For
the AR(1) model of Eq. \ref{eq:xprime}, the PSD turns out to be \cite{priestley}
\begin{equation}
S\left(f\right)=\frac{\alpha\left(2-\alpha\right)}{1-2\left(1-\alpha\right)\cos\left(2\pi f\right)+\left(1-\alpha\right)^{2}}\sigma_{x}^{2}.\label{eq:sfMod}
\end{equation}
This is again analogous to an OU process, since the Fourier transform
of an exponential function is a Lorentzian function. There are significant
oscillations only in the case $\alpha\lesssim2$, for which the PSD
peaks at high frequencies (Fig. \ref{fig:oscillations}cd). The case
of \emph{E. coli}, where $\alpha\approx1/2$,\emph{ }is far from this
regime (Fig. \ref{fig:oscillations}ab). Therefore, experimentally
observed fluctuations should not be confused for oscillations \cite{you}.

The AR(1) model in Eq. \ref{eq:xprime} can be extended to incorporate
details that are relevant to a variety of microorganisms. These include
asymmetric and noisy divisions (e.g. in mycobacteria \cite{aldridge,logsdon}),
noisy growth rates (e.g. in slow growing \emph{E. coli }\cite{elf}
and in the archaeon \emph{H. salinarum} \cite{eun}), and diverse
growth morphologies (e.g. the budding mode of growth of \emph{S. cerevisiae}
\cite{soifer}). First, noisy divisions and noisy growth rates can
be incorporated by modeling the division ratio (daughter cell size
at birth divided by mother cell size at division) and growth rate
as $1/2+\xi_{r}$ and $\lambda+\xi_{\lambda}$, respectively, at each
generation. If the fluctuations $\xi_{r}$ and $\xi_{\lambda}$ are
small and uncorrelated, Eq. \ref{eq:xprime} becomes to first order
in small variables
\begin{equation}
x'\approx\left(1-\alpha\right)x+\lambda\xi_{t}+2\xi_{r}.\label{eq:extended}
\end{equation}
The fluctuation $\xi_{\lambda}$ enters as a first order correction
to interdivision time $t_{d}$. The CVs and CCs can be calculated
as before for Eq. \ref{eq:extended} to show that the different fluctuations
typically affect the CVs and CCs in different ways. For example, the
CC between cell size at birth and at division, $C\left(v_{b},v_{d}\right)$,
is sensitive to fluctuations in division ratios, and is increased
by large fluctuations in division ratios. However, the CC in cell
size at birth between mother-daughter pairs, $C\left(v_{b},v_{b}'\right)$,
remains the same as in Eq. \ref{eq:cvbvd}, and is independent of
all noise terms. It is thus a robust detector of the underlying regulation
strategy even in face of multiple sources of complicating stochasticity
\cite{eun}. We discuss later several models that incorporate other
biological details and move beyond AR models.

\subsection{Continuous rate models and higher order effects\label{subsec:CRM}}

Cell size regulation can also be modeled using continuous rate models
(CRM) \cite{jun,lagomarsinoPRE16,lagomarsinoPRE17,lagomarsinoConcerted}.
In contrast to DSMs, CRMs consider not just discrete division events,
but the continuous cell cycle. They specify the instantaneous division
rate $h$, or the probability to divide per unit size increment, as
a function of physiological parameters such as the current size $v$,
size at birth $v_{b}$, growth rate $\lambda$, or the time $t$ since
division. A simple choice of parametrization is the sloppy sizer model,
$h=h\left(v\right)$. In this case, the probability for a cell of
size $v$ to divide between the size interval $v$ and $v+dv$ is
$h\left(v\right)dv$. Hence if $F\left(v_{d}|v_{b}\right)$ is the
cumulative probability to have not divided at size $v_{d}$ given
$v_{b}$, then $F\left(v_{d}|v_{b}\right)$ satisfies $F\left(v_{d}+dv|v_{b}\right)=F\left(v_{d}|v_{b}\right)\left(1-h\left(v_{d}\right)dv\right)$.
In the continuum limit, $dF\left(v_{d}|v_{b}\right)/dv=-h\left(v_{d}\right)F\left(v_{d}|v_{b}\right)$,
so that
\begin{equation}
F\left(v_{d}|v_{b}\right)=\exp\left(-\int_{v_{b}}^{v_{d}}h\left(v\right)dv\right).\label{eq:cumpvdvb}
\end{equation}
Eq. \ref{eq:cumpvdvb} can be written as
\begin{equation}
h\left(v\right)=-\frac{d}{dv}\ln F\left(v|v_{b}\right),\label{eq:hdv}
\end{equation}
allowing to extract $h\left(v\right)$ via single-cell experiments
that measure $F\left(v_{d}|v_{b}\right)$. The division rate can be
formulated as a probability to divide per unit time increment as well,
using the change of variables between size and time given by exponential
growth. Analyses using CRMs have demonstrated that the current size
is not the only determinant of the division rate because the sloppy
sizer model fails to capture measured distributions of interdivision
time and size increment from birth to division \cite{lagomarsinoConcerted}.
This implies that there exists a feedback on the time since birth,
or equivalently the size at birth \cite{lagomarsinoConcerted}. Specifically,
a division rate in the form $h=h\left(v-v_{b}\right)$ can simultaneously
describe measured distributions of size at birth, interdivision time,
and size increment from birth to division \cite{jun}.

A CRM can be approximately reduced to a DSM with the target size at
division equal to the expectation value of the size at division given
the size at birth \cite{lagomarsinoPRE17},
\begin{equation}
f\left(v_{b}\right)=\int_{0}^{\infty}p\left(v|v_{b}\right)vdv,\label{eq:fvbpv}
\end{equation}
where $p\left(v|v_{b}\right)=-dF\left(v|v_{b}\right)/dv$ is the probability
density for a cell born at size $v_{b}$ to divide at size $v$. The
nature and magnitude of the noise term can be determined by inverting
the steps described below to map a DSM to a corresponding CRM. To
do so, $p\left(v|v_{b}\right)$ can be calculated from $f\left(v_{b}\right)$
and a specified coarse-grained noise, then the division rate can be
obtained using Eq. \ref{eq:hdv}. For example, for a time-additive,
normally distributed noise with variance $\sigma_{t}^{2}$, the division
probability density for log-size $x=\ln\left(v/v_{0}\right)$ is $p\left(x|x_{b}\right)\propto\exp\left(-\left(x-g\left(x_{b}\right)\right)^{2}/\left(2\lambda^{2}\sigma_{t}^{2}\right)\right)$,
where $g\left(x_{b}\right)=\ln\left(f\left(v_{0}e^{x_{b}}\right)/v_{0}\right)$.
Since the typical $x_{b}$ is much smaller than $g\left(x_{b}\right)$,
integration leads to the division rate \cite{lagomarsinoPRE17}
\begin{equation}
h\left(v,v_{b}\right)\approx\frac{\sqrt{2}}{v\sqrt{\pi}\lambda\sigma_{t}}H\left(\frac{1}{\sqrt{2}\lambda\sigma_{t}}\ln\left(\frac{v}{f\left(v_{b}\right)}\right)\right),\label{eq:h}
\end{equation}
where $H\left(z\right)=\exp\left(-z^{2}\right)/\left(1-\textnormal{Erf}\left(z\right)\right)$
and $\textnormal{Erf}\left(\cdot\right)$ is the error function. For
the regulatory function Eq. \ref{eq:fchoice}, the division rate Eq.
\ref{eq:h} becomes a function of only the instantaneous size when
$\alpha=1$, corresponding to a sizer strategy.

Although the CRM is generic and may capture complex behavior such
as filamentation \cite{lagomarsinoConcerted}, it is not obvious a
priori how to parametrize the division rate. On the other hand, the
DSM has only a few parameters, is amenable to analytical treatment
in several cases, and describes existing measurements well. The complexity
sacrificed by DSMs and their first order approximate solutions may
become important, for instance, when second and higher order terms
become significant. However, higher order effects are difficult to
detect unless the number of cells measured is large enough to suppress
the confounding effects of fluctuations in the cell cycle. No existing
experiments have achieved this regime, perhaps justifying the success
of DSMs as models of cell size regulation \cite{lagomarsinoPRE17}.

\subsection{More precise analyses of DSMs}

The approximate first order solution in Section \ref{subsec:firstorder}
predicts a log-normal size distribution for a regulatory function
$f\left(v_{b}\right)=2\left(1-\alpha\right)v_{b}+v_{0}$ and a time-additive,
normally distributed noise. However, closer inspection reveals that
the size distribution has a power-law tail instead. This can be seen
by analyzing which moments exist for a given regulation strength $\alpha$.
Calculations similar to those in Section \ref{subsec:firstorder}
show that if the $j$-th moment exists, so too do all the lower moments,
but that for any $\alpha>0$, there always exists an integer $j^{*}$
past which all moments cease to exist. This suggests that the size
distribution has a power-law tail $p\left(v_{b}\right)\sim1/v_{b}^{1+\beta}$
with $j^{*}<\beta\leq j^{*}+1$, as confirmed by numerical simulations
\cite{marantan}.

The value of $\beta$ can be obtained precisely. The evolution of
size distributions from one generation to the next can be written
as an integral equation $p\left(v'_{b}\right)=\int_{0}^{\infty}K\left(v'_{b},v_{b}\right)p\left(v_{b}\right)dv_{b},$
where the kernel $K\left(v_{b}',v_{b}\right)$ can be derived from
the regulatory strategy $f\left(v_{b}\right)$. For a regulatory function
in the form $f\left(v_{b}\right)=2\left(1-\alpha\right)v_{0}\left(v_{b}/v_{0}\right)^{\eta}+v_{0}$,
it can be shown via an asymptotic analysis of the integral equation
that a distribution with a power-law tail $1/v_{b}^{1+\beta}$ evolves
to one with a power-law tail $1/v_{b}^{1+\beta/\eta^{2}}$ \cite{marantan}.
This implies that for $\eta=1$, the stable size distribution indeed
has a power-law tail, with
\begin{equation}
\beta=\frac{-2\ln\left(1-\alpha\right)}{\lambda^{2}\sigma_{t}^{2}},\label{eq:beta}
\end{equation}
where $\sigma_{t}^{2}$ is the variance of the time-additive noise.

An alternative approach also led to the same power-law tail \cite{burov}.
In this approach, a DSM is approximated as a Langevin equation continuous
in generations. Let $x=\ln\left(v_{b}/v_{0}\right)$ be the log-size
at birth and let $n$ denote the generation number, then Eq. \ref{eq:DSM}
can be written
\begin{equation}
x_{n+1}=x_{n}+\tilde{g}\left(x_{n}\right)+\lambda\xi_{t},\label{eq:discrete}
\end{equation}
where $\tilde{g}\left(x_{n}\right)=\ln\left(f\left(\exp\left(x_{n}\right)v_{0}\right)/v_{0}\right)-x_{n}$.
To lowest order, Eq. \ref{eq:discrete} can be approximated by a Langevin
equation continuous in $n$ as
\begin{equation}
\frac{dx}{dn}=\tilde{g}\left(x\right)+\lambda\xi_{t}.\label{eq:continuousn}
\end{equation}
As seen before in the context of an OU process described by Eq. \ref{eq:ou},
Eq. \ref{eq:continuousn} leads to an equilibrium distribution of
log-sizes $p\left(x\right)\propto\exp\left(-2V\left(x\right)/\left(\lambda^{2}\sigma_{t}^{2}\right)\right)$,
where $V\left(x\right)=\int\tilde{g}\left(x'\right)dx'$ is the effective
potential. For the same regulatory function as above, the effective
potential diverges linearly as $V\left(x\right)\sim-2x\ln\left(1-\alpha\right)$.
The equilibrium distribution therefore has a power-law tail $1/v_{b}^{1+\beta}$
with the same $\beta$ as in Eq. \ref{eq:beta} \cite{burov}. Even
further precision can be obtained via a second order approximation
which modifies the effective potential, but leaves the behavior of
the power-tail unchanged \cite{burov}.

\section{Beyond phenomenological models of cell size regulation}

As we previously alluded, the formalism of DSMs developed for the
problem of cell size regulation can lead to insights on related problems
at the molecular, single-cell, and the population level. Below, we
discuss these in turn.

\subsection{Molecular mechanisms to implement cell size regulation\label{subsec:molmech}}

How does a bacterial cell molecularly implement a size regulation
strategy? The initiator accumulation model is a network architecture
proposing that an initiator protein accumulates during cell growth
to trigger cell division upon reaching a threshold copy number $\theta$
\cite{sompayrac}. While experiments have suggested that the upstream
control occurs over initiation of DNA replication rather than cell
division in various microorganisms \cite{donachie,liu,soifer,amirSpandrel},
we first review a simpler model where the accumulation of initiators
triggers cell division. The model leads to the adder correlations
observed in several species of bacteria and other microorganisms \cite{barber,amir,ho}.

One possible molecular implementation of the initiator accumulation
model is as follows \cite{singh}. If the transcription rate of the
initiator is assumed to be proportional to the cell volume, which
grows exponentially in time, and if each transcript leads to a burst
of protein production with mean burst size $b$, then the distribution
of added cell size $\Delta v=v_{d}-v_{b}$ from birth to division
has width \cite{singh}
\begin{equation}
CV^{2}\left(\Delta v\right)=\frac{b^{2}+2b\theta+\theta}{\left(b+\theta\right)^{2}}.
\end{equation}
Furthermore, the resulting distribution has only one characteristic
size, the mean added cell size $\left\langle \Delta v\right\rangle $,
and therefore can be written as
\begin{equation}
p\left(\Delta v\right)=\frac{1}{\left\langle \Delta v\right\rangle }\tilde{p}\left(\frac{\Delta v}{\left\langle \Delta v\right\rangle }\right).
\end{equation}
Indeed, experiments showed that distributions of cell sizes with different
means collapse after normalizing by the mean \cite{jun,lagomarsinoPRE16,woldringh,giometto}.
The collapse suggests that $b$ and $\theta$ are constant within
the implementation here. The same experiments also saw that the distributions
of interdivision times collapse after normalizing by the mean doubling
time, which is again captured by this model \cite{singh}. There are
also additional models that show such scaling collapse, such as an
autocatalytic network subject to a threshold criterion for division
\cite{iyerbiswas}, and the coarse-grained ``adder-per-origin''
model described below.

As discussed in the Introduction, control at other cell cycle events
may lie upstream of cell division in various microorganisms. DSMs
similar to those reviewed so far can be extended to describe cell
cycle regulation. These models can not only produce emergent strategies
of cell size regulation identical to those described by the division-centric
models reviewed so far, but also describe additional statistics such
as the correlations between cell size and various cell cycle timings
\cite{lagomarsinoTrends,adicipt}.

As an example, we review below a model of cell cycle regulation in
\emph{E. coli}, whose cell divisions appear to follow a constant time
$T$ after the initiation of DNA replication for a broad range of
mean growth rates \cite{huang,ch}. The time $T$ can be larger than
the mean doubling time $\tau$, in which case the cells maintain multiple
ongoing rounds of DNA replication. The tight coupling between initiation
and division implies that the cell size at birth $v_{d}$ is
\begin{equation}
v_{d}=v_{i}e^{\lambda\left(T+\xi_{T}\right)},\label{eq:ch}
\end{equation}
where $v_{i}$ is the cell size at initiation, $\lambda=\ln2/\tau$
is the growth rate, and $\xi_{T}$ describes fluctuations with magnitude
$\sigma_{T}$ in the time between initiation and division. At a coarse-grained
level, the initiator accumulation model can be described as \cite{sompayrac,amir,ho}
\begin{equation}
\tilde{v}_{i}'=\left(v_{i}+Ov_{0}\right)e^{\lambda\xi_{t}},\label{eq:apo}
\end{equation}
where $\tilde{v}_{i}'$ is the total cell size of the daughter cells
(typically two) at the next initiation, $O$ is the number of origins
of replication (i.e. the site along the chromosome at which DNA replication
initiates), and $v_{0}$ is a constant. As in the division-centric
model, regulation is subject to a time-additive noise $\xi_{t}$ with
magnitude $\sigma_{t}$.

Analysis and simulations of the initiation-centric model of Eqs. \ref{eq:ch}-\ref{eq:apo}
show that it produces emergent adder correlations at division \cite{ho,barber},
as long as the magnitude of the fluctuations in the coordination between
initiation and division is much less than that in the control of initiation
($\sigma_{t}\gg\sigma_{T}$). This is indeed the case in experiments
for fast-growing bacteria, although the picture appears different
for slow-growing bacteria \cite{elf}, which we discuss later. The
model also generates cell size and interdivision time distributions
whose CVs only depend on the magnitudes $\lambda\sigma_{t}$ and $\lambda\sigma_{T}$
of the fluctuations, and the regulation strength $\alpha$. The distributions
therefore collapse after scaling by the mean if these parameters are
constant across growth conditions. At the bulk-level, the initiation-centric
model produces the observed exponential scaling of mean cell size
with mean growth rate, as discussed in the Introduction, without requiring
parameters to depend on mean growth rate \cite{ho}. These results,
together with previous results regarding the universality of cell
size distributions, suggest that the initiator accumulation model
may be a robust molecular mechanism that produces adder correlations,
and that models of cell cycle regulation can continue to shed light
on the underlying biology.

\subsection{Regulation of protein numbers}

Recent works have begun investigating the statistics of the copy numbers
of proteins at the single-cell level in the same spirit as the problem
of cell size regulation \cite{you,brennerEPJE,brennerPRE,brennerMulticomp}.
In fact, for a constitutively expressed protein, the distributions
of protein numbers at birth can be described by a DSM \cite{brennerPRE}.
Analysis analogous to that in Fig. \ref{fig:analysis}, but for the
copy number of a constitutively expressed protein in \emph{E. coli}
in the same data set \cite{youData}, reveals that the partial CC
is also zero in this case. However, it is unclear how protein number
and cell size are simultaneously regulated.

One way to investigate this question is via a multi-dimensional, or
vector, AR model. An AR(1) vector model in $M$ dimensions can be
written as

\begin{equation}
\vec{x}'=A\vec{x}+\vec{b}+\vec{\xi},\label{eq:multiD}
\end{equation}
where $\vec{x}$ is a vector of the abundances at birth of the $M$
cellular components, which can include cell size, and $\vec{x}'$
is the vector at the next generation. $A$ is a $M\times M$ matrix
representing the regulatory interactions between components, $\vec{b}$
is a vector representing the basal synthesis level between generations,
and $\vec{\xi}$ is a vector of noise terms uncorrelated between generations
but may be cross-correlated at the same generation.

For the one-dimensional case, the condition for stationarity is that
$2>\alpha>0$ so that the variance of $x$ in Eq. \ref{eq:sigx2}
is finite. This condition is equivalent to that the zero of $1-\left(1-\alpha\right)z$
lie outside the unit circle. In the multi-dimensional case, the condition
is similarly that all the zeros of $\det\left(I-Az\right)$ lie outside
the unit circle \cite{priestley}. Given a stable AR(1) vector model,
the multi-dimensional analogue of the Yule-Walker equations can be
used to estimate by maximum likelihood the regulatory matrix $A$
from measurements \cite{box}. For the data set discussed above \cite{youData},
this method\emph{ }results in
\[
A=\left(\begin{array}{cc}
0.50\pm0.02 & -0.02\pm0.01\\
-0.16\pm0.02 & 0.60\pm0.02
\end{array}\right),
\]
where the first and second components are respectively cell size and
protein number at birth (both normalized by their means), and plus-minus
shows the standard error in the estimate. This novel result suggests
that the copy number of this constitutively expressed protein does
not affect cell size regulation, while cell size does affect the regulation
of this protein number. It is unknown whether this result holds for
all constitutively expressed proteins, and how this result will change
for proteins that are not constitutively expressed.

To better understand cross-correlations between cell size and protein
numbers from a mechanistic perspective, recent works have investigated
a dynamical model in the form $d\vec{x}/dt=A\vec{x}$, where $\vec{x}$
is now the abundances of the cellular components during the cell cycle,
and $A$ now describes the regulatory interactions in time \cite{brennerMulticomp}.
This model leads to the components growing as a sum of exponentials
that can be approximated as a single exponential function during one
generation, in agreement with experimentally observed exponential
growth \cite{brennerEPJE}. Describing the statistics generated by
dynamical models, and relating a dynamical model to a DSM and vice
versa remain important open questions.

\subsection{Effects of cell size regulation on population growth rate}

\begin{figure*}
\noindent \begin{centering}
\includegraphics[width=6in]{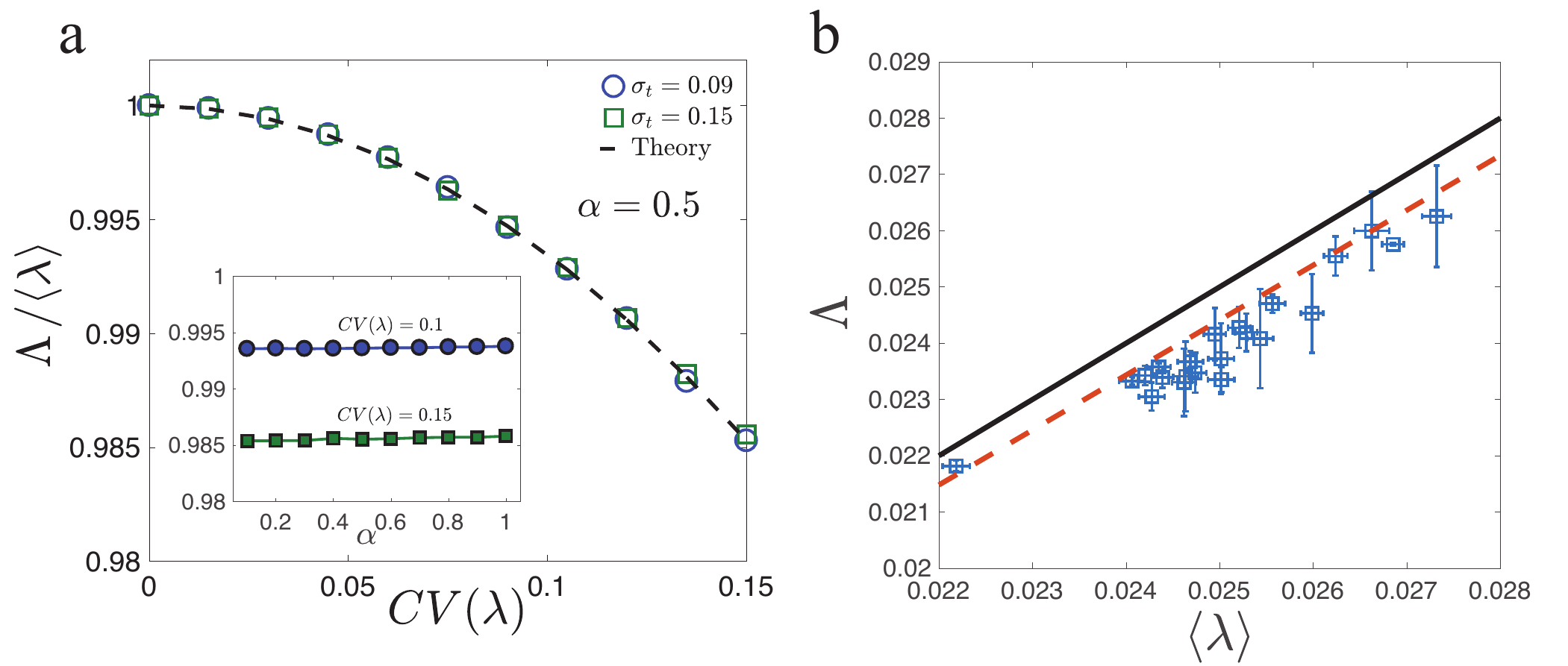}
\par\end{centering}
\caption{\label{fig:population} Cell size regulation, as long as it exists,
does not affect population growth rate. (a) Population growth rates
obtained from simulations (symbols) of an exponentially growing population
subject to variability in single-cell growth rates agree with Eq.
\ref{eq:poplambda} (dashed line). Inset shows population growth rates
do not vary with the regulation strength as long as $1>\alpha>0$.
(b) Variability in single-cell growth rates decreases population growth
rate. Blue squares show data from \cite{stewart} (details of the
error bars can be found in \cite{lin}). Red dashed line shows Eq.
\ref{eq:poplambda} for $CV\left(\lambda\right)$ measured by experiments.
Black solid line shows $\Lambda=\left\langle \lambda\right\rangle $
as a guide. Both axes have units $\textnormal{min}^{-1}$. Adapted
from \cite{lin}.}
\end{figure*}

At the single-cell level, genetically identical cells in the same
clonal populations may have different interdivision times and growth
rates. How does such variability at the single-cell level affect population
growth? Models often assume that the interdivision times $t_{d}$
are uncorrelated between generations and independent of other variables
\cite{powell,hashimoto,iyerbiswasPop}. In this case, a simple relation
connects the asymptotic population growth rate $\Lambda=\left(dN/dt\right)/N$,
where $N$ is the number of cells in the population, to the interdivision
time distribution $p\left(t_{d}\right)$,
\begin{equation}
2\int_{0}^{\infty}p\left(t_{d}\right)\exp\left(-\Lambda t_{d}\right)=1.\label{eq:popClassic}
\end{equation}
Importantly, given a fixed mean interdivision time, a larger variability
in $t_{d}$ increases $\Lambda$. However, cell size regulation leads
to negative correlations in $t_{d}$ between generations, as seen
in Eq. \ref{eq:ctdtd}. In this case, recent results obtained by some
of us showed that in an asynchronous, exponentially growing population
- in which each cell is subject to variability in its single-cell
growth rate, as well as to time-additive and size-additive noise in
its cell size regulation by the regulatory function $f\left(v_{b}\right)=2\left(1-\alpha\right)v_{b}+2\alpha v_{0}$
- the population growth rate is dependent only on the distribution
of single-cell growth rates. In the limit of small correlations in
growth rates between generations, variability in single-cell growth
rates does not increase, but rather decreases the population growth
rate \cite{lin},
\begin{equation}
\Lambda/\left\langle \lambda\right\rangle =1-\left(1-\frac{\ln2}{2}\right)CV^{2}\left(\lambda\right),\label{eq:poplambda}
\end{equation}
Eq. \ref{eq:poplambda} predicts that a population can enhance its
population growth rate by suppressing the variability in single-cell
growth rates given a fixed mean, which is consistent with the smaller
CV of single-cell growth rates than that of interdivision times observed
in experiments (Fig. \ref{fig:population}ab) \cite{jun,elf}. Eq.
\ref{eq:poplambda} holds for any size regulation strategy $1>\alpha>0$,
implying that cell size regulation, as long as it exists (in particular,
$\alpha\neq0$ leads instead to Eq. \ref{eq:popClassic}), does not
affect population growth rate within the models studied here (Fig.
\ref{fig:population}a).

\section{Discussion}

In this review, we summarized the mathematical formulations of the
problem of cell size regulation, with a focus on coarse-grained, discrete
models. As an example, we showed that a first order autoregressive
model can describe the statistics of cell size in \emph{E. coli}.
We discussed how detailed analyses of such models led to several biologically
relevant insights at the molecular, single-cell, and population level.
The same approach may shed light on several outstanding questions.

First, the prevalence of adder correlations in all three domains of
life (e.g. the prokaryote \emph{E. coli} \cite{jun,jw,elf}, the eukaryote
\emph{S. cerevisiae} \cite{soifer}, and the archaeon \emph{H. salinarum}
\cite{eun}) suggest that it may be simpler to implement or may be
evolutionarily advantageous compared to other size regulation strategies.
An explanation of the prevalence of adder correlations remains missing,
however, since cell size regulation was found not to affect population
growth rate within the class of phenomenological growth models reviewed
here \cite{lin}.

In contrast, the mean single-cell growth rate affects cell size regulation,
since \emph{E. coli} in slow growth conditions no longer exhibits
adder correlations \cite{elf}. This observation may be explained
by introducing stochasticity in single-cell growth rates into the
initiation-centric model discussed in Section \ref{subsec:molmech},
which can lead to a size regulation strategy that varies with mean
growth rates \cite{elf}. Indeed, cell size regulation may potentially
be an emergent property of cell cycle regulation \cite{amirSpandrel}.
This view is further supported by several models that describe cell
division as a downstream effect of another cell cycle event (e.g.
initiation of DNA replication in \emph{M. smegmatis }\cite{logsdon},
the onset of budding in \emph{S. cerevisiae} \cite{barber,soifer},
and septum constriction in \emph{C. crescentus }\cite{dinner}), and
nonetheless reproduces the observed statistics at divisions.

Models of cell size regulation may also incorporate diverse growth
morphologies. For example in \emph{S. cerevisiae}, division asymmetry
depends on the duration of the budded phase, during which all cell
growth occurs for the budded daughter cell \cite{soifer}. In \emph{M.
smegmatis}, cell growth occurs at the two poles of the cell: the old
pole grows faster than the new pole, and on average, the daughter
that inherits the old pole is larger \cite{aldridge}. In both cases,
the subpopulations formed by the larger and smaller daughter cells
exhibit different emergent size regulation strategies \cite{logsdon,soifer}.
Models incorporating these details move beyond AR models but remain
straightforward to simulate numerically, allowing the statistics they
generate to be compared to experiments to distinguish between competing
models.

At the molecular level, the behavior of molecular network architectures
require further analysis. The particular implementation of an initiator
accumulation model discussed in Section \ref{subsec:molmech} made
the strong assumption that transcription rate is proportional to cell
volume \cite{singh}. It would be interesting to study more detailed
network architectures, where this would be a result rather than a
model assumption. Models at the molecular level may also begin to
investigate the problem of protein number regulation. Since proteins
are made by ribosomes, an important problem is how ribosomes are allocated
towards translating different types of proteins. Quantitative patterns
at the bulk level have emerged regarding ribosome allocation \cite{hwa,barkai},
but the picture at the dynamical, cell cycle level is less clear.
Models of stochastic gene expression that extend existing ones, which
often consider a fixed cell volume \cite{paulsson}, to incorporate
cell cycle regulation could shed light in this aspect.

Incorporating cell cycle regulation can in turn help understand cell
size regulation in organisms with a circadian clock, such as mammalian
cells and cyanobacteria. Recent works have begun to examine cell size
regulation in these organisms \cite{balaban,balaban2,huangCyano,martins},
and some have suggested that the circadian clock may affect cell size
regulation in these organisms. How to model these processes at a molecular,
coarse-grained, and population level remain intriguing questions.

\subsection{Which model should we use?}

Since the various classes of models of cell size regulation reviewed
here are quite different fundamentally, we conclude by discussing
the question of which model to use to analyze what data or to elucidate
what phenomenon. First, as discussed in Section \ref{subsec:CRM},
continuous rate models (CRMs) differ from discrete stochastic maps
(DSMs) in that CRMs take as parameter an entire function that describes
the instantaneous division rate, whereas DSMs take, for example, only
the strength of regulation and the magnitude of the coarse-grained
stochasticity (although the form of the stochasticity must be assumed).
Moreover, CRMs assume a priori whether regulation depends only on
the current size or also on the size at birth, whereas DSMs can be
used to determine the mode of regulation. These could be reasons for
the increasing visibility of DSMs as models of cell size regulation.

Another fundamental distinction is the difference between division-centric
models and models that place control at an upstream event. Division-centric
models, such as those described by Eq. \ref{eq:DSM}, makes the strong
assumption that all information relevant for determining division
timing is stored in the current size and the size at birth. This does
not have to be the case. For example, it is widely accepted that in
\emph{S. cerevisiae}, control occurs over the Start transition and
the duration of the budded phase is uncorrelated with size \cite{soifer}.
As reviewed briefly in Section \ref{subsec:molmech}, Eq. \ref{eq:DSM}
can be adapted to place control at various cell cycle events, which
may then lead to additional predictions that can illuminate the coupling
between different cell cycle events.

The above distinction does not imply that we should always use the
most detailed description. To cite Levins, ``All models leave out
a lot and are in that sense false, incomplete, inadequate. The validation
of a model is not that it is 'true' but that it generates good testable
hypotheses relevant to important problems \cite{levins}.'' It is
often helpful to sacrifice details not pertinent to the phenomenon
under consideration. For example, the initiator accumulation model
can be considered to trigger division rather than initiation. Yet
the simplified model may still provide mechanistic insights into the
statistical properties, as discussed in Section \ref{subsec:molmech}.
These mechanistic models are altogether different from the phenomenological
DSMs and CRMs.

In summary, the appropriate model to use depends not only on the organism
or system in question, but also on the phenomena explored within the
model. We have sketched here a map of the existing models. Technological
advances now enable collection of more accurate and larger data sets.
These will likely stimulate further development of models, which in
turn will influence experimental directions.

%\begin{summary}[SUMMARY POINTS]
\section*{Summary points}
\begin{enumerate}
\item Studying quantitative patterns associated with cell size, and modeling
them using stochastic models, can shed light on the underlying biological
mechanisms.
\item Cell size in \emph{E. coli} can be described by a first order autoregressive
model in which the present value depends only on the value in the
previous generation.
\item The initiator accumulation model is a molecular network architecture
of cell size regulation that appears to be consistent with existing
experimental results in bacteria.
\item Cell size regulation, as long as it is present, does not affect population
growth rate within existing models.
\end{enumerate}
%\end{summary}

%\begin{issues}[FUTURE ISSUES]
\section*{Future issues}
\begin{enumerate}
\item What is the reason for the prevalence of adder correlations for cell
size regulation?
\item What are molecular implementations that regulate the cell cycle in
changing environments, and what are the limits of biological stochasticity
they can sustain?
\item How are the copy number of proteins and cell size simultaneously regulated,
and how can the resulting statistics be described?
\item What are the couplings between cell size and cell cycle regulation
to other cellular processes such as circadian clocks, and what models
should we use to describe them?
\end{enumerate}
%\end{issues}

\section*{Acknowledgments}

This manuscript has been submitted to the Annual Review of Biophysics.
We thank Lingchong You for the permission to use the data from Ref.
\cite{youData}. PH was supported by the Harvard MRSEC program of
the National Science Foundation under award number DMR 14-20570. JL
was supported by the Carrier Fellowship. AA thank the Alfred P. Sloan
Foundation Research Fellowship, Harvard Dean\textquoteright s Competitive
Fund for Promising Scholarship, the Milton Fund, the Kavli Foundation,
and the Volkswagen Foundation for their support.


\begin{thebibliography}{00}

\bibitem{koch}
A~Koch.
\newblock {\em Bacterial growth and form}.
\newblock Springer, 2001.

\bibitem{huang}
L~Willis and KC~Huang.
\newblock Sizing up the bacterial cell cycle.
\newblock {\em Nat. Rev. Microbiol.}, 15(10):606--20, 2017.

\bibitem{smk}
M~Schaechter, O~Maaloe, and NO~Kjeldgaard.
\newblock Dependency on medium and temperature of cell size and chemical
  composition during balanced growth of \emph{Salmonella typhimurium}.
\newblock {\em J. Gen. Microbiol.}, 19:592--606, 1958.

\bibitem{levin}
NS~Hill, R~Kadoya, DK~Chattoraj, and PA~Levin.
\newblock Cell size and the initiation of dna replication in bacteria.
\newblock {\em PLos Genet.}, 8(3):e1002549, 2012.

\bibitem{jun}
S~Taheri-Araghi, S~Bradde, JT~Sauls, NS~Hill, PA~Levin, J~Paulsson,
  M~Vergassola, and S~Jun.
\newblock Cell-size control and homeostasis in bacteria.
\newblock {\em Curr. Biol.}, 25(3):385--91, 2015.

\bibitem{donachie}
WD~Donachie.
\newblock Relationship between cell size and time of initiation of dna
  replication.
\newblock {\em Nature}, 219(5158):1077--79, 1968.

\bibitem{theriot}
L~Harris and J~Theriot.
\newblock Relative rates of surface and volume synthesis set bacterial cell
  size.
\newblock {\em Cell}, 165(6):1479--1492, 2016.

\bibitem{jw}
M~Campos, IV~Surotsev, S~Kato, A~Paintdakhi, B~Beltran, SE~Ebmeier, and
  C~Jacobs-Wagner.
\newblock A constant size extension drives bacterial cell size homeostasis.
\newblock {\em Cell}, 159(6):1433--46, 2014.

\bibitem{liu}
H~Zheng, P~Ho, M~Jiang, B~Tang, W~Liu, D~Li, X~Yu, NE~Kleckner, A~Amir, and
  C~Liu.
\newblock Interrogating the \emph{Escherichia coli} cell cycle by cell
  dimension perturbations.
\newblock {\em Proc. Natl. Acad. Sci.}, 113(52):15000--5, 2016.

\bibitem{godin}
M~Godin, FF~Delgado, S~Son, WH~Grover, AK~Bryan, A~Tzur, P~Jorgensen, K~Payer,
  AD~Grossman, MW~Kirschner, and SR~Manalis.
\newblock Using buoyant mass to measure the growth of single cells.
\newblock {\em Nat. Methods}, 7(5):387--90, 2010.

\bibitem{sompayrac}
L~Sompayrac and O~Maaloe.
\newblock Autorepressor model for control of dna replication.
\newblock {\em Nat. New. Biol.}, 241(109):133--5, 1973.

\bibitem{fantes}
PA~Fantes.
\newblock Control of cell size and cycle time in \emph{Schizosaccharomyces
  pombe}.
\newblock {\em J. Cell. Sci.}, 24:51--67, 1977.

\bibitem{barber}
F~Barber, P~Ho, A~Murray, and A~Amir.
\newblock Details matter: noise and model structure set the relationship
  between cell size and cell cycle timing.
\newblock {\em Front. Cell Dev. Biol.}, in press.

\bibitem{marantan}
A~Marantan and A~Amir.
\newblock Stochastic modeling of cell growth with symmetric or asymmetric
  division.
\newblock {\em Phys. Rev. E}, 94:012405, 2016.

\bibitem{boxQuote}
GEP Box, JS~Hunter, and WG~Hunter.
\newblock {\em Statistics for Experimenters}.
\newblock John Wiley and Sons, 2005.

\bibitem{gardiner}
C~Gardiner.
\newblock {\em Stochastic methods: A handbook for the natural and social
  sciences}.
\newblock Springer, 2009.

\bibitem{youData}
Y~Tanouchi, A~Pai, H~Park, S~Huang, NE~Buchler, and L~You.
\newblock Long-term growth data of \emph{Escherichia coli} at a single-cell
  level.
\newblock {\em Sci Data}, 4:170036, 2017.

\bibitem{junWang}
P~Wang, L~Robert, J~Pelletier, WL~Dang, F~Taddei, A~Wright, and S~Jun.
\newblock Robust growth of \emph{Escherichia coli}.
\newblock {\em Curr. Biol.}, 20(12):1099--103, 2010.

\bibitem{taheriaraghiRev}
S~Taheri-Araghi, SD~Brown, JT~Sauls, DB~McIntosh, and S~Jun.
\newblock Single-cell physiology.
\newblock {\em Annu. Rev. Biophys.}, 44:123--42, 2015.

\bibitem{amir}
A~Amir.
\newblock Cell size regulation in bacteria.
\newblock {\em Phys. Rev. Lett.}, 112(20):208102, 2014.

\bibitem{elf}
M~Wallden, D~Fange, EG~Lundius, O~Baltekin, and J~Elf.
\newblock The synchronization of replication and division cycles in individual
  \emph{E. coli} cells.
\newblock {\em Cell}, 166(3):729--39, 2016.

\bibitem{junReview}
JT~Sauls, D~Li, and S~Jun.
\newblock Adder and a coarse-grained approach to cell size homeostasis in
  bacteria.
\newblock {\em Curr. Opin. Cell Biol.}, 38:38--44, 2016.

\bibitem{voorn}
WJ~Voorn, LJ~Koppes, and NB~Grover.
\newblock Mathematics of cell division in \emph{Escherichia coli}: comparison
  between sloppy-size and incremental-size kinetics.
\newblock {\em Curr. Top. Mol. Gen.}, 1:187--194, 1993.

\bibitem{koppes}
WJ~Voorn and LJ~Koppes.
\newblock Skew or third moment of bacterial generation times.
\newblock {\em Arch. Microbiol.}, 169(1):43--51, 1997.

\bibitem{soifer}
I~Soifer, L~Robert, and A~Amir.
\newblock Single-cell analysis of growth in budding yeast and bacteria reveals
  a common size regulation strategy.
\newblock {\em Curr. Biol.}, 26(3):356--361, 2016.

\bibitem{koppes80}
LJ~Koppes, M~Meyer, HB~Oonk, MA~de~Jong, and N~Nanninga.
\newblock Correlation between size and age at different events in the cell
  division cycle of \emph{Escherichia coli}.
\newblock {\em J. Bacteriol.}, 143(3):1241--52, 1980.

\bibitem{lin}
J~Lin and A~Amir.
\newblock The effects of stochasticity at the single-cell level and cell size
  control on the population growth.
\newblock {\em Cell Systems}, 5:1--10, 2017.

\bibitem{priestley}
MB~Priestley.
\newblock {\em Spectral analysis and time series}.
\newblock Academic Press, 1981.

\bibitem{balaban}
O~Sandler, S~Pearl-Mizrahi, N~Weiss, O~Agam, I~Simon, and NQ~Balaban.
\newblock Lineage correlations of single cell division time as a probe of
  cell-cycle dynamics.
\newblock {\em Nature}, 519(7544):468--71, 2015.

\bibitem{balaban2}
N~Mosheiff, BMC Martins, S~Pearl-Mizrahi, A~Gruenberger, S~Helfrich,
  I~Mihalcescu, D~Kohlheyer, JCW Locke, L~Glass, and NQ~Balaban.
\newblock Correlations of single-cell division times with and without periodic
  forcing.
\newblock {\em arXiv}, page 1710.00349, 2017.

\bibitem{you}
Y~Tanouchi, A~Pai, H~Park, S~Huang, R~Stamatov, N~Buchler, and L~You.
\newblock A noisy linear map underlies oscillations in cell size and gene
  expression in bacteria.
\newblock {\em Nature}, 523(7560):357--60, 2015.

\bibitem{aldridge}
BB~Aldridge, M~Fernandez-Suarez, D~Heller, V~Ambravaneswaran, D~Irimia,
  M~Toner, and SM~Fortune.
\newblock Asymmetry and aging of mycobacterial cells lead to variable growth
  and antibiotic susceptibility.
\newblock {\em Science}, 335(6064):100--4, 2012.

\bibitem{logsdon}
MM~Logsdon, P~Ho, K~Papavinasasundaram, M~Cokol, K~Richardson, CM~Sassetti,
  A~Amir, and BB~Aldridge.
\newblock Coordination of cell cycle progression in mycobacteria.
\newblock {\em Curr. Biol.}, in press.

\bibitem{eun}
Y~Eun, P~Ho, M~Kim, L~Renner, S~LaRussa, L~Robert, A~Schmid, E~Garner, and
  A~Amir.
\newblock Archaeal cells share common size control with bacteria despite
  noisier growth and division.
\newblock under review.

\bibitem{lagomarsinoPRE16}
AS~Kennard, M~Osella, A~Javer, J~Grilli, P~Nghe, S~Tans, P~Cicuta, and
  MC~Lagomarsino.
\newblock Individuality and universality in the growth-division laws of single
  \emph{E. coli} cells.
\newblock {\em Phys. Rev. E}, 93:012408, 2016.

\bibitem{lagomarsinoPRE17}
J~Grilli, M~Osella, AS~Kennard, and MC~Lagomarsino.
\newblock Relevant parameters in models of cell division control.
\newblock {\em Phys. Rev. E}, 95(3):032411, 2017.

\bibitem{lagomarsinoConcerted}
M~Osella, E~Nugent, and MC~Lagomarsino.
\newblock Concerted control of \emph{Escherichia coli} cell division.
\newblock {\em Proc. Natl. Acad. Sci.}, 111(9):3431--5, 2014.

\bibitem{burov}
DA~Kessler and S~Burov.
\newblock Effective potential for cellular size control.
\newblock {\em arXiv}, page 1701.01725, 2017.

\bibitem{amirSpandrel}
A~Amir.
\newblock Is cell size a spandrel?
\newblock {\em eLife}, 6:e22186, 2017.

\bibitem{ho}
P~Ho and A~Amir.
\newblock Simultaneous regulation of cell size and chromosome replication in
  bacteria.
\newblock {\em Front. Microbiol.}, 6:662, 2015.

\bibitem{singh}
KR~Ghusinga, C~Vargas-Garcia, and A~Singh.
\newblock A mechanistic stochastic framework for regulating bacterial cell
  division.
\newblock {\em Sci. Rep.}, 6:30229, 2016.

\bibitem{woldringh}
FJ~Trueba, OM~Neijssel, and CL~Woldringh.
\newblock Generality of the growth kinetics of the average individual cell in
  different bacterial populations.
\newblock {\em J. Bacteriol.}, 150(3):1048--55, 1982.

\bibitem{giometto}
A~Giometto, F~Altermatt, F~Carrara, A~Maritan, and A~Rinaldo.
\newblock Scaling body size fluctuations.
\newblock {\em Proc. Natl. Acad. Sci.}, 110(12):4646--50, 2012.

\bibitem{iyerbiswas}
S~Iyer-Biswas, GE~Crooks, NF~Scherer, and AR~Dinner.
\newblock Universality in stochastic exponential growth.
\newblock {\em Phys. Rev. Lett.}, 113:028101, 2014.

\bibitem{lagomarsinoTrends}
M~Osella, SJ~Tans, and MC~Lagomarsino.
\newblock Step by step, cell by cell: quantification of the bacterial cell
  cycle.
\newblock {\em Trends. Microbiol.}, 25(4):250--6, 2017.

\bibitem{adicipt}
A~Adiciptaningrum, M~Osella, MC~Moolman, MC~Lagomarsino, and S~Tans.
\newblock Stochasticity and homeostasis in the \emph{E. coli} replication and
  division cycle.
\newblock {\em Sci. Rep.}, 5:18261, 2015.

\bibitem{ch}
S~Cooper and CE~Helmstetter.
\newblock Chromosome replication and the division cycle of \emph{Escherichia
  coli} b/r.
\newblock {\em J. Mol. Biol.}, 31(3):519--40, 1968.

\bibitem{brennerEPJE}
N~Brenner, E~Braun, A~Yoney, L~Susman, J~Rotella, and H~Salman.
\newblock Single-cell protein dynamics reproduce universal fluctuations in cell
  populations.
\newblock {\em Eur. Phys. J. E}, 38:102, 2015.

\bibitem{brennerPRE}
N~Brenner, CM~Newman, D~Osmanovic, Y~Rabin, H~Salman, and DL~Stein.
\newblock Universal protein distributions in a model of cell growth and
  division.
\newblock {\em Phys. Rev. E}, 92(4):042713, 2015.

\bibitem{brennerMulticomp}
L~Susman, M~Kohram, H~Vashistha, JT~Nechleba, H~Salman, and N~Brenner.
\newblock Statistical properties and dynamics of phenotype components in
  individual bacteria.
\newblock {\em arXiv}, page 1609.05513, 2017.

\bibitem{box}
GEP Box, GM~Jenkins, and GC~Reinsel.
\newblock {\em Time series analysis: Forecasting and control}.
\newblock Prentice Hall, 1994.

\bibitem{powell}
EO~Powell.
\newblock Growth rate and generation time of bacteria, with special reference
  to continuous culture.
\newblock {\em J. Gen. Microbiol.}, 15:492--511, 1956.

\bibitem{hashimoto}
M~Hashimoto, T~Nozoe, H~Nakaoka, R~Okura, S~Akiyoshi, K~Kaneko, E~Kussell, and
  Y~Wakamoto.
\newblock Noise-driven growth rate gain in clonal cellular populations.
\newblock {\em Proc. Natl. Acad. Sci.}, 113(12):3251--6, 2016.

\bibitem{iyerbiswasPop}
S~Iyer-Biswas, H~Gudjonson, CS~Wright, J~Riebling, E~Dawson, K~Lo, A~Fiebig,
  S~Crosson, and AR~Dinner.
\newblock Bridging the time scales of single-cell and population dynamics.
\newblock {\em arXiv}, page 1611.05149, 2016.

\bibitem{stewart}
EJ~Stewart, R~Madden, G~Paul, and F~Taddei.
\newblock Aging and death in an organism that reproduces by morphologically
  symmetric division.
\newblock {\em PLoS Biol.}, 3(2):e45, 2005.

\bibitem{dinner}
S~Banerjee, K~Lo, MK~Daddysman, A~Selewa, T~Kuntz, AR~Dinner, and NF~Scherer.
\newblock Biphasic growth dynamics control cell division in \emph{Caulobacter
  crescentus}.
\newblock {\em Nat. Microbiol.}, 2:17116, 2017.

\bibitem{hwa}
M~Scott, CW~Gunderson, EM~Mateescu, Z~Zhang, and T~Hwa.
\newblock Interdependence of cell growth and gene expression: origins and
  consequences.
\newblock {\em Science}, 330(6007):1099--102, 2010.

\bibitem{barkai}
E~Metzl-Raz, M~Kafri, G~Yaakov, I~Soifer, Y~Gurvich, and N~Barkai.
\newblock Principles of cellular resource allocation revealed by
  condition-dependent proteome profiling.
\newblock {\em eLife}, 6:e28034, 2017.

\bibitem{paulsson}
J~Paulsson.
\newblock Models of stochastic gene expression.
\newblock {\em Phys. Life Rev.}, 2:157--175, 2005.

\bibitem{huangCyano}
FB~Yu, L~Willis, RMW Chau, A~Zambon, M~Horowitz, D~Bhaya, KC~Huang, and
  SR~Quake.
\newblock Long-term microfluidic tracking of coccoid cyanobacterial cells
  reveals robust control of division timing.
\newblock {\em BMC Biol.}, 15:11, 2017.

\bibitem{martins}
BMC Martins, AK~Tooke, P~Thomas, and JCW Locke.
\newblock Cell size control driven by the circadian clock and environment in
  cyanobacteria.
\newblock {\em bioRxiv}, page 183558, 2017.

\bibitem{levins}
R~Levins.
\newblock The strategy of model building in population biology.
\newblock {\em Amer. Sci.}, 54(4):421--31, 1966.

\end{thebibliography}
\end{document}